\pdfoutput=1

\documentclass[11pt,twoside,a4paper,cmspaper,final,collab]{cms-tdr}
\def\svnVersion{124535}\def\cmsCernNo{2012-057}\def\cmsCernDate{\today}\def\cmsMessage{Submitted to the Journal of High Energy Physics}

\begin{document}\cmsNoteHeader{BPH-10-015}

\hyphenation{had-ron-i-za-tion}
\hyphenation{cal-or-i-me-ter}
\hyphenation{de-vices}
\RCS$Revision: 116898 $
\RCS$HeadURL: svn+ssh://alverson@svn.cern.ch/reps/tdr2/papers/BPH-10-015/trunk/BPH-10-015.tex $
\RCS$Id: BPH-10-015.tex 116898 2012-04-18 17:28:25Z mgalanti $
\newcommand{\ups}{$\Upsilon$ }

\newcommand{\ptjpsi}{{\ensuremath{p_{T}({J/\psi})}}}
\newcommand{\etajpsi}{{\ensuremath{y({J/\psi})}}}
\newcommand{\ptmu}{{\ensuremath{p_\mathrm{T}^{\mu}}}}
\newcommand{\etamu}{{\ensuremath{\eta^{\mu}}}}
\newcommand{\dsigmadpt}{{\ensuremath{\mathrm{d}\sigma/\\mathrm{d}p_\mathem{T}}}}
\newcommand{\ptb}{{\ensuremath{p_\mathrm{T}^{H_\cPqb}}}}

\newcommand{\effic}{{\ensuremath{\epsilon}}}
\cmsNoteHeader{BPH-10-015}
\title{Measurement of the inclusive cross section \texorpdfstring{$\sigma(\Pp\Pp\rightarrow \bbbar \mbox{X}\rightarrow \Pgm\Pgm \mbox{X}^\prime)$ at $\sqrt{s}=7$ TeV}{
for production of b b-bar X decaying to muons in pp collisions at sqrt(s)=7 TeV}}

\date{\today}

\abstract{
A measurement of the inclusive cross section for the process
$ \Pp\Pp \rightarrow \bbbar \mbox{X} \rightarrow \Pgm\Pgm \mbox{X}^\prime$
at $\sqrt{s} = 7\TeV$ is presented,
based on a data sample corresponding to an integrated luminosity of 27.9\pbinv collected by the
CMS experiment at the LHC.
By selecting pairs of muons each with pseudorapidity $|\eta|<2.1$,
the value
$\sigma ( \Pp\Pp \rightarrow \bbbar\mbox{X}
\rightarrow \Pgm\Pgm \mbox{X}^\prime)=26.4\pm0.1 \mbox{ (stat.) }\pm2.4\mbox{ (syst.) }\pm1.1\mbox{ (lumi.) }$\,nb
is obtained for muons with transverse momentum $\pt>4$\GeV, and
$5.12\pm0.03 \mbox{ (stat.) }\pm0.48\mbox{ (syst.) }\pm0.20\mbox{
  (lumi.)}\unit{nb}$ for $\pt>6$\GeV.
These results are compared to QCD predictions at leading and
next-to-leading orders.}

\hypersetup{%
pdfauthor={CMS Collaboration},%
pdftitle={Measurement of the cross section for production of b b-bar X, decaying to muons in pp collisions at sqrt(s)= 7 TeV},%
pdfsubject={CMS},%
pdfkeywords={CMS, physics, software, computing}}

\maketitle 

\clearpage
\section{Introduction}

The measurement of the cross section for inclusive $\cPqb$-quark production at
the Large Hadron Collider (LHC) is a powerful probe of quantum chromodynamics (QCD) at very high energies.
In addition, knowledge of the inclusive $\cPqb$-production rate from QCD
processes helps to understand the background in searches for massive
particles decaying into $\cPqb$ quarks, such as the Higgs
boson or new heavy particles.

The $\cPqb$-quark production cross section can be computed at next-to-leading order (NLO) in a
perturbative QCD expansion~\cite{Nason:1987xz,Nason:1989zy,Beenakker:1990maa}.
 The sizeable scale dependence of the result
suggests that the contribution from the neglected
higher-order terms is large~\cite{Collins:1991ty,Catani:1990eg,Cacciari:1993mq}.
The measurements performed at the Tevatron in $\Pp\Pap$ collisions
at $\sqrt{s}=1.8$ and $1.96\TeV$~\cite{Abbott:1999se,CDF-B-BBAR},
and at the LHC by the Compact Muon Solenoid
(CMS)~\cite{BPH_10_004,BPH_10_005,BPH_10_007} and
LHCb~\cite{LHCb_b_bbar_cs,LHCb_bplus_cs} collaborations in $\Pp\Pp$ collisions at $\sqrt{s}=7\TeV$ in different
rapidity ranges are generally consistent with the theoretical calculations. However, the comparisons are affected by large
theoretical uncertainties.

The measurements of the cross section for the inclusive process
$ \Pp\Pp\rightarrow \bbbar \mbox{X} \rightarrow \Pgm \Pgm \mbox{X}^\prime$
at $\sqrt{s}=7\TeV$ presented here allow  for a comparison with QCD predictions
 in a kinematic domain where NLO calculations are more
reliable because of the suppressed contribution of the gluon-splitting production mechanism
 (as discussed in \cite{Frixione1997} and the references therein).

Experimentally, the dimuon final state allows for the selection
of a sample with high $\bbbar$ event purity
in the following wide kinematical region:
muon pseudorapidity $|\eta|<2.1$, where
$\eta = - \ln \left[ \tan \left( \theta/2 \right) \right]$
and  $\theta$ is the angle between the muon momentum
and the counterclockwise beam direction,
and muon momentum in the
plane transverse to the beam axis $\pt>4\GeV$ or $\pt>6\GeV$.
Discrimination of the background from charm and light quark decays and from
the Drell--Yan process
is accomplished using the two-dimensional distribution of the two muon impact
parameters ($d_{xy}$), defined as the distance of closest
approach of each muon track to the interaction point projected onto the
plane transverse to the beam axis.

This paper is structured as follows. A brief description of the CMS
detector is presented in Section \ref{s:CMS}. Section \ref{s:data}
describes the collision and simulated data used for this measurement and
the selection criteria.
Section \ref{s:sample} contains a detailed description of the
categories in which events are grouped according to each muon's
production process and kinematic features, while
the fit to the impact parameter distributions is discussed in Section \ref{s:fit}.
Section \ref{s:eff} describes how the efficiency is computed and
Section \ref{s:syst} is devoted to the determination of the
systematic uncertainties.
Section \ref{s:cross_section} reports the cross section measured in data and expected from QCD predictions.

\section{The CMS detector}\label{s:CMS}

A detailed description of the CMS experiment can be found elsewhere~\cite{JINST}.
The central feature of the CMS apparatus is a superconducting 3.8\unit{T}
solenoid of 6\unit{m} internal diameter.  Within the field volume are the
silicon tracker, the crystal electromagnetic
calorimeter (ECAL), and the brass/scintillator hadron calorimeter
(HCAL).
Muons are detected in the pseudorapidity range $|\eta|< 2.4$
by gaseous detectors utilizing three technologies: drift tubes (DT), cathode
strip chambers (CSC), and resistive plate chambers (RPC), embedded in the steel
return yoke.
The silicon tracker is composed of pixel detectors (three barrel
layers and two forward disks on either side of the detector, made of
66 million $100\mum\times150\mum$ pixels) followed by microstrip detectors
(ten barrel layers, three inner disks and nine forward disks on either side of the
detector, with the strip pitch between 80 and 180\mum).
Thanks to the strong magnetic field and
high granularity of the silicon tracker, the
transverse momentum $\pt$ of muons matched to reconstructed
tracks is measured with the resolution better than $1.5\%$  for
$\pt <$ 100\GeV.
 The silicon tracker also provides the vertex position with
 ${\sim}15\mum$ accuracy.
The impact parameter
resolution is measured
with a sample of muons
from
$\PgUa \rightarrow \Pgmp\Pgmm$ decays to be 28\mum and 21\mum for
muons with $\pt>4\GeV$ and $\pt>6\GeV$, respectively.

The first level (L1) of the CMS trigger system, composed of custom
hardware processors, uses information from the calorimeters and muon
detectors to select the most interesting events.
The rapidity coverage of the L1 muon triggers used in this analysis is $|\eta|<2.4$.
The high-level-trigger
 processor farm further decreases the event rate before data storage.

\section{Data selection and Monte Carlo simulation}\label{s:data}

The data employed for this measurement were collected with the CMS
detector during the 2010 running period of the LHC. They correspond to
an integrated luminosity $\mathcal{L} = 27.9\pm 1.1\pbinv$~\cite{lumipas}.
A sample of events with two muons, each with transverse momentum $\pt>3\GeV$
were selected at the trigger level.
Further requirements, designed to increase the purity
of the muon candidates and
to increase the fraction of muons from $\cPqb$ decay in the sample,
are applied
at the analysis stage.
 A muon candidate is selected by matching information
from the silicon tracker and muon chambers.
The track must contain at least 12 hits from the silicon
tracker, with signals in at least two pixel layers,
and a normalized $\chi^2$ not exceeding 2.
The overall $\chi^2$ obtained
by combining the information from the tracker and
the muon chambers should not exceed 10 times the number
of degrees of freedom. Finally, each muon must be contained in the
kinematical region defined by
$|\eta| < 2.1$ and
$\pt>4\GeV$.
We perform the measurement in this region and in a higher $\pt$ region
where both the muons have
$\pt>6\GeV$.

Primary interaction vertices are reconstructed event-by-event from
the reconstructed tracks. A candidate vertex
is accepted if its fit has at least four degrees of freedom and its distance
from the beam spot does not exceed 24\unit{cm} along the beam line and 1.8\unit{cm}
in the plane transverse to the beams.
Tracks are assigned to
the primary vertex for which the track's distance to the vertex along
the beam direction is smallest at the point of closest approach in the
transverse plane.
Muon tracks are required to have an impact
parameter $d_{xy}$ perpendicular to the beam direction and with respect to its assigned primary vertex
of less than 0.2\,cm.
Events are kept only if both
muon tracks are assigned to the same primary vertex and both cross the beam axis
within 1 cm of that vertex position along the beam direction.

To remove muons from $\PZz$ decays, a selection on the dimuon mass
$M_{\mu\mu} <70\GeV$ is applied.
The mass range contributed by the $\PgU$ resonances, $8.9\GeV <M_{\Pgm\Pgm} < 10.6\GeV$, is also rejected.
Charmonium resonances and sequential semileptonic decays from a single
\cPqb\ quark
(for example
$\cPqb \rightarrow \JPsi~\mbox{X} \rightarrow\Pgm\Pgm \mbox{X}$, or
$\cPqb \rightarrow \cPqc\Pgm\mbox{X} \rightarrow \Pgm\Pgm \mbox{X}^\prime$)
are rejected
by removing dimuons with $M_{\Pgm\Pgm} < 5\GeV$.
Events are selected if one and only one pair of muons is found
satisfying all the criteria defined above. A total of
537\,734 events for $\pt > 4\GeV$ and 151\,314 events for $\pt > 6\GeV$
pass these requirements.

Two samples of simulated Monte Carlo (MC) events were generated using the minimum-bias settings
of \PYTHIA 6.422 \cite{pythia64} (parameter MSEL=1),
with the Z2 tune \cite{MCtunes,QCD-10-010}, and incorporating the CTEQ6L1 parton distribution functions (PDF) \cite{cteqPaper}.
To increase the generation efficiency within the selected acceptance,
a filter was applied at the generator level requiring two muons with
$\pt^{\text{gen}}>2.5\GeV$ and $|\eta^{\text{gen}}|<2.5$ for the measurement with $\pt > 4\GeV$, or
$\pt^{\text{gen}}>5\GeV$ and $|\eta^{\text{gen}}|<2.5$ for the measurement with $\pt > 6\GeV$.
The generated samples include events
with muons originating from the decay of light
mesons (mostly charged pions and kaons) within the tracker volume.
A third MC sample was produced
to simulate the Drell--Yan process.
MC events, including the full simulation of the CMS detector and trigger via the \GEANTfour
package~\cite{geantfour}, are subjected to the same
reconstruction and selection as the real data.

\section{Templates for different muon classes\label{s:sample}}

 The fraction
 of signal events
 ($\Pp\Pp \rightarrow \cPqb\cPaqb \mbox{X} \rightarrow \Pgm\Pgm \mbox{X}^\prime$)
 in the data
 is obtained from a fit to the
2D distribution of the impact parameters
 of the two muons.
 For this purpose,
 reconstructed muons
 in the simulated events
 are separated into
 four different classes, defined according to
 their origin.
 The single-particle distributions of the transverse impact parameter $d_{xy}$ are obtained
 for each class from simulation and
fit
using analytical functions.
 From these functions, the 2D templates are built symmetrically.
 This procedure is described in the following section.

\subsection{Definition of muon classes}\label{s:muon_classes}

Information from the generation process is used to assign each reconstructed
muon in the simulation to a well-defined category. Reconstructed muon
candidates are linked to the corresponding generated charged
particles with a hit-based associator, which
reduces the probability of incorrect
associations to a negligible level.
Tracks are assigned to one of the following classes:

\begin{enumerate}
\item B-hadron decays (B):
muons produced in the decay of a $B$ hadron, including both
  direct decays ($\cPqb\rightarrow \Pgmm \mbox{X}$) and
  cascade decays
  ($\cPqb\rightarrow \cPqc\mbox{X} \rightarrow \Pgm \mbox{X}^\prime$,
$\cPqb\rightarrow \Pgt\mbox{X} \rightarrow \Pgm \mbox{X}^\prime$,
$\cPqb\rightarrow \JPsi~\mbox{X} \rightarrow\Pgm^\pm \mbox{X}$);
\item Charmed hadron decays (C):
muons from the semileptonic decays of charmed hadrons produced
promptly;
\item Prompt tracks (P): candidates originating from the primary
  vertex, mostly  muons
  from the Drell--Yan process and quarkonia decays.
  This category also includes punch through of primary
  hadrons, and muons from decays of charged pions and kaons in the
  volume between the silicon tracker and the muon chambers;
\item Decays in flight (D):
muons produced in decays of
 charged pions or kaons (which may come either from light- or
 heavy-flavor hadrons) in the silicon tracker volume.
\end{enumerate}

Table \ref{t:sample} gives the single-muon sample composition from the
simulation for MC events passing the full selection and dimuon trigger.
While the fraction of muons from decays in flight (D) decreases at
larger $\pt$, the prompt component (P) increases due to the Drell--Yan muons.

The predicted
composition of the dimuon events from the simulation is shown in Table~\ref{t:sample2D},
where PX is defined as the sum of the PB, PC, and PD contributions.
The uncertainties given in the table are the statistical uncertainties from
the simulated samples.
\begin{table}[h!]
  \centering
\caption{Percentage of each muon class in the simulated
  events for two \pt\ requirements. The uncertainties are statistical only.}
\label{t:sample}
  \begin{tabular}{l|c|c}
    \hline
 Source  & \multicolumn{2}{c}{Fraction in simulation (\%)}                         \\
                 &   $\pt>4\GeV$           &       $\pt>6\GeV$       \\
    \hline
    B hadron (B)          &    $77.8\pm0.2 $      &   $  79.8\pm 0.4$       \\
    C hadron (C)         &   $  14.0\pm0.1$      &   $  12.6\pm 0.1$       \\
    Prompt sources (P)   &   $    1.84\pm 0.04$     &    $   3.44\pm 0.08$      \\
    Decays in flight (D)  &   $    6.37\pm 0.07$     &    $   4.21\pm 0.09$    \\
    \hline
\end{tabular}
\end{table}

\begin{table}[h!]
\centering
\caption{Percentage of dimuon event sources in the
  simulation for two different \pt\ requirements. PX represents the sum of the contributions from PB, PC, and PD. The uncertainties are statistical only.}
\label{t:sample2D}
\begin{tabular}{c|c|c}
\hline
Source  & \multicolumn{2}{c}{Fraction in simulation (\%)}                         \\
                 &   $\pt>4\GeV$           &       $\pt>6\GeV$       \\
\hline
BB             &      $71.6 \pm 0.2$      &       $74.6 \pm 0.4$      \\
CC            &       $9.24 \pm 0.08$       &       $8.67 \pm 0.14$       \\
BC            &       $5.66 \pm 0.07$       &        $5.22 \pm 0.11$      \\
PP             &       $1.84 \pm 0.04$       &        $3.43 \pm 0.08$              \\
DD           &       $1.49 \pm 0.04$       &        $0.73 \pm 0.04$      \\
BD            &       $6.01 \pm 0.07$       &        $4.40 \pm 0.10$      \\
CD           &       $3.69 \pm 0.05$       &        $2.53 \pm 0.08$ \\
PX            &       $0.48 \pm 0.02$       &        $0.40 \pm 0.03$ \\
\hline
\end{tabular}
\end{table}

\begin{figure}[!htb]
  \begin{center}
    \includegraphics[height=0.45\textwidth, angle=0]{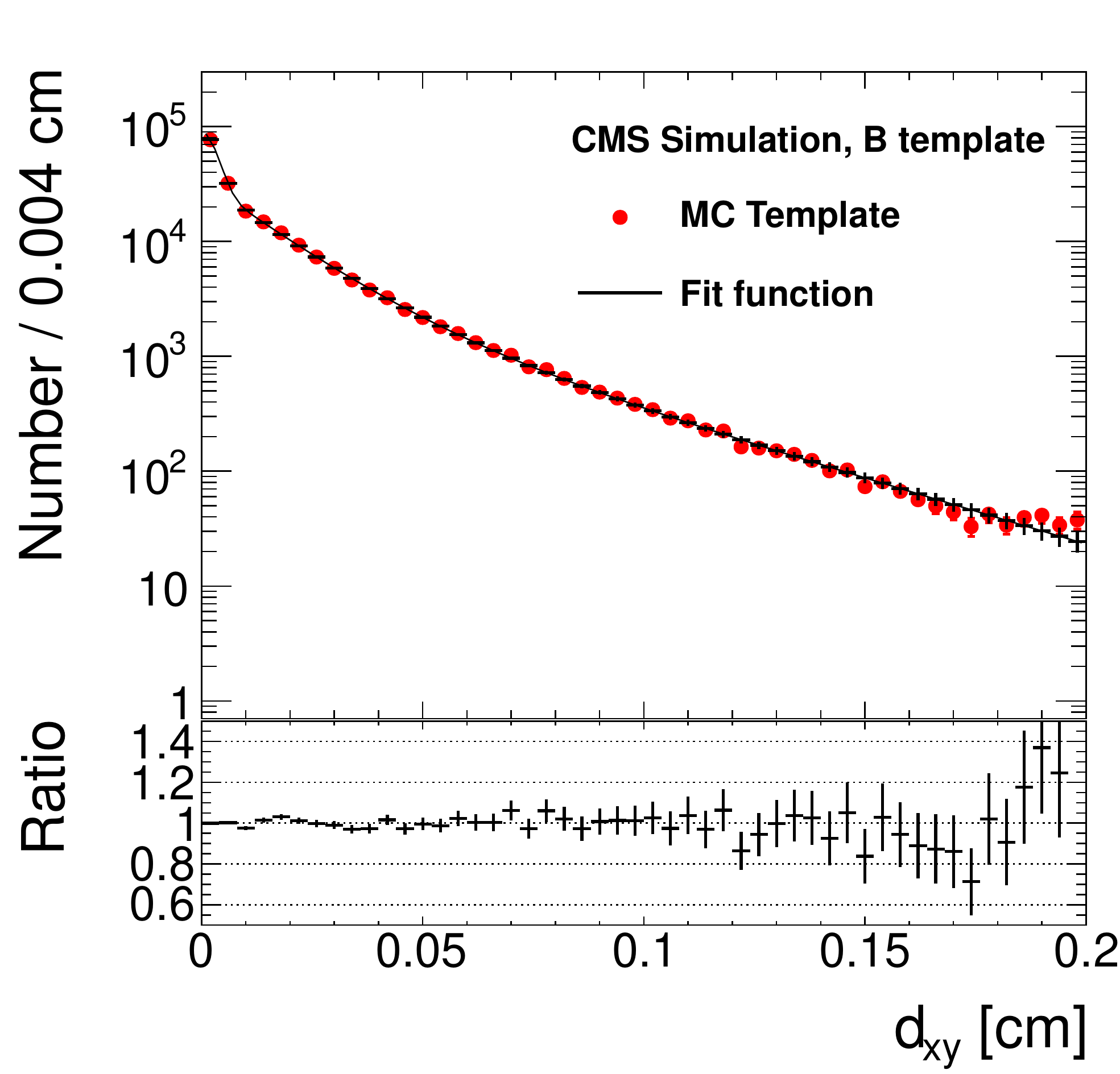}
    \includegraphics[height=0.45\textwidth, angle=0]{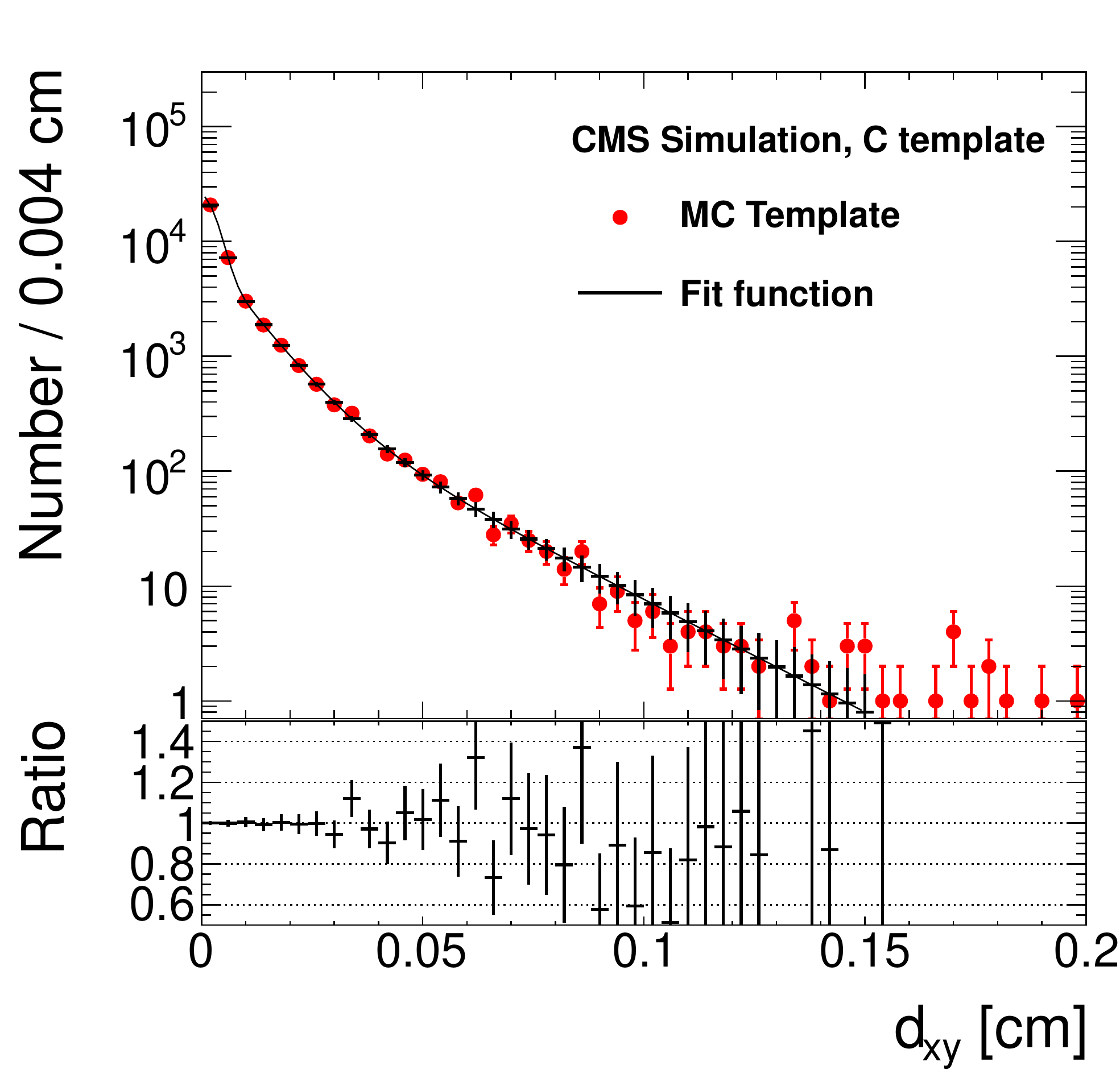}
    \includegraphics[height=0.45\textwidth, angle=0]{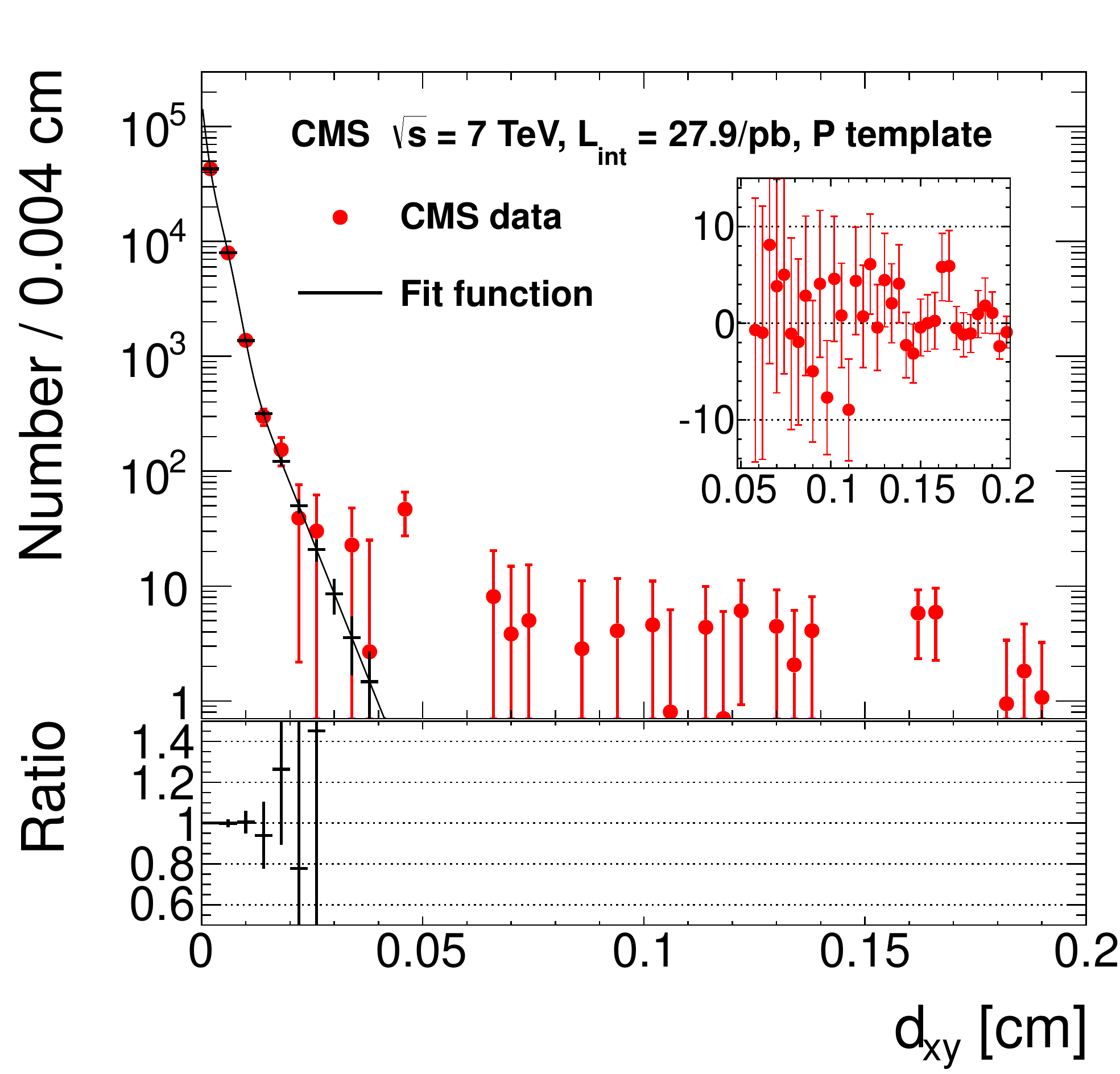}
    \includegraphics[height=0.45\textwidth, angle=0]{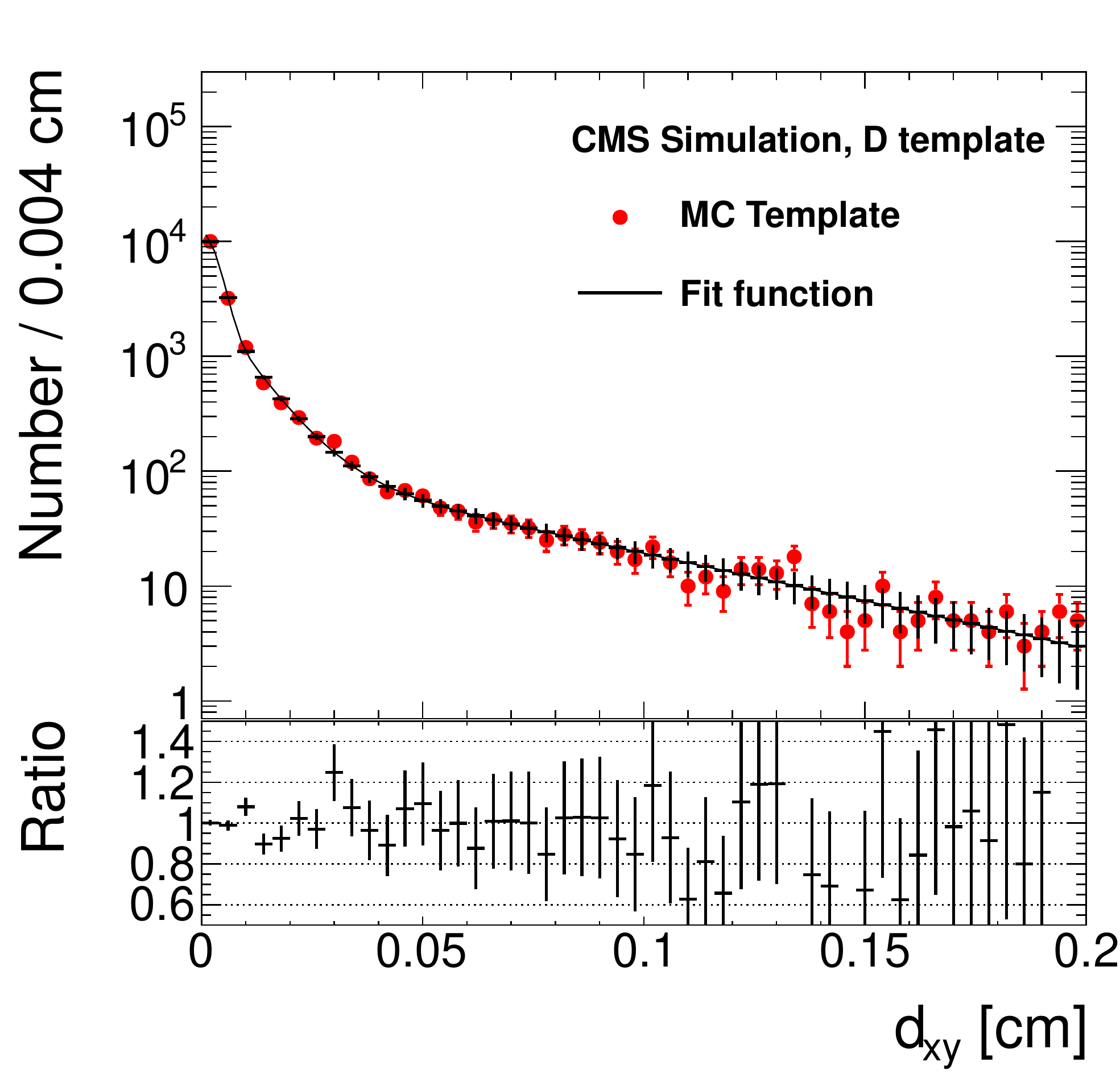}

    \caption{Comparison, for muons with $\pt>4\GeV$, between the template $d_{xy}$ histogram (red)
             and the fitted function (black) for muons coming from
    B hadrons
    (class B, top left),
charmed hadrons
(class C, top right),
prompt tracks (class P, bottom left), and decays in flight (class D, bottom right).
The templates for B, C, and D come from simulation.
For the prompt tracks, the distribution is obtained from data.
An enlargement of the prompt-track distribution for $d_{xy}>0.05\cm$
is shown on a linear scale as an insert in the lower-left plot.
For each template, the ratio of the $d_{xy}$
histogram to the fitted function is shown at the bottom.}
          \label{f:1Dtemplate_b_c_p_d}
      \end{center}
    \end{figure}

Figure~\ref{f:1Dtemplate_b_c_p_d} shows the $d_{xy}$ distributions
 for muons with $\pt>4\GeV$ from
the simulation for all the classes above
except for the prompt tracks, where muons from
decays of $\PgUa$ in the collision data are used
after removing the background with a sideband subtraction technique.

The prompt $d_{xy}$ distribution is fit with the sum of a Gaussian centered at zero
 and an exponential function.
This combination of functions accounts for the detector resolution effects.
The distributions of the other classes are fit
using, in addition, a second exponential term.
The functions are shown by continuous black lines overlaid on the
histograms in Fig.~\ref{f:1Dtemplate_b_c_p_d}, while the black points
represent the template histograms obtained by evaluating the fit
functions at each bin center. The ratio of the MC distribution to the
fit values are shown in the lower plots of Fig.~\ref{f:1Dtemplate_b_c_p_d}.
The templates for muons with $\pt>6\GeV$ are obtained in a similar
way.

\subsection{Two-dimensional template distributions}

In principle, the dimuon events could be split
into sixteen different categories by combining the four classes defined above
for each muon.
In order to reduce the number of categories to ten,
the $d_{xy}$ distributions are symmetrized (i.e., BC=CB, BD=DB, etc.)
using a method
originally developed by the CDF collaboration \cite{CDF-B-BBAR}.
The one-dimensional (1D) histograms, built as described above, are
normalized to unity within the fit range
$0<d_{xy}<0.2\,\mbox{cm}$.
The symmetrized
2D template histogram for the events with a muon of class
$\rho$ and another of class $\sigma$
($\rho,\,\sigma =1,\ldots,4$ according to the definition in
Section \ref{s:muon_classes})
is then constructed as:
\begin{linenomath}
\begin{equation}
  T^{\rho,\sigma}_{ij}=\frac{1}{2} (S^{\rho}_{i} S^{\sigma}_{j}+ S^{\rho}_{j} S^{\sigma}_{i}),
\end{equation}
\end{linenomath}
where $S^{\rho}_i$ is the content of the $i^{\text{th}}$ bin of the
histogram describing the class $\rho$, and analogously for index $j$ and
class $\sigma$.
In this way, ten symmetric distributions are obtained. In
practice, the few events from the PX category are
neglected, thus reducing the number of significant classes to seven.

\begin{figure}[!htb]
      \begin{center}
           \includegraphics[height=0.45\textwidth]{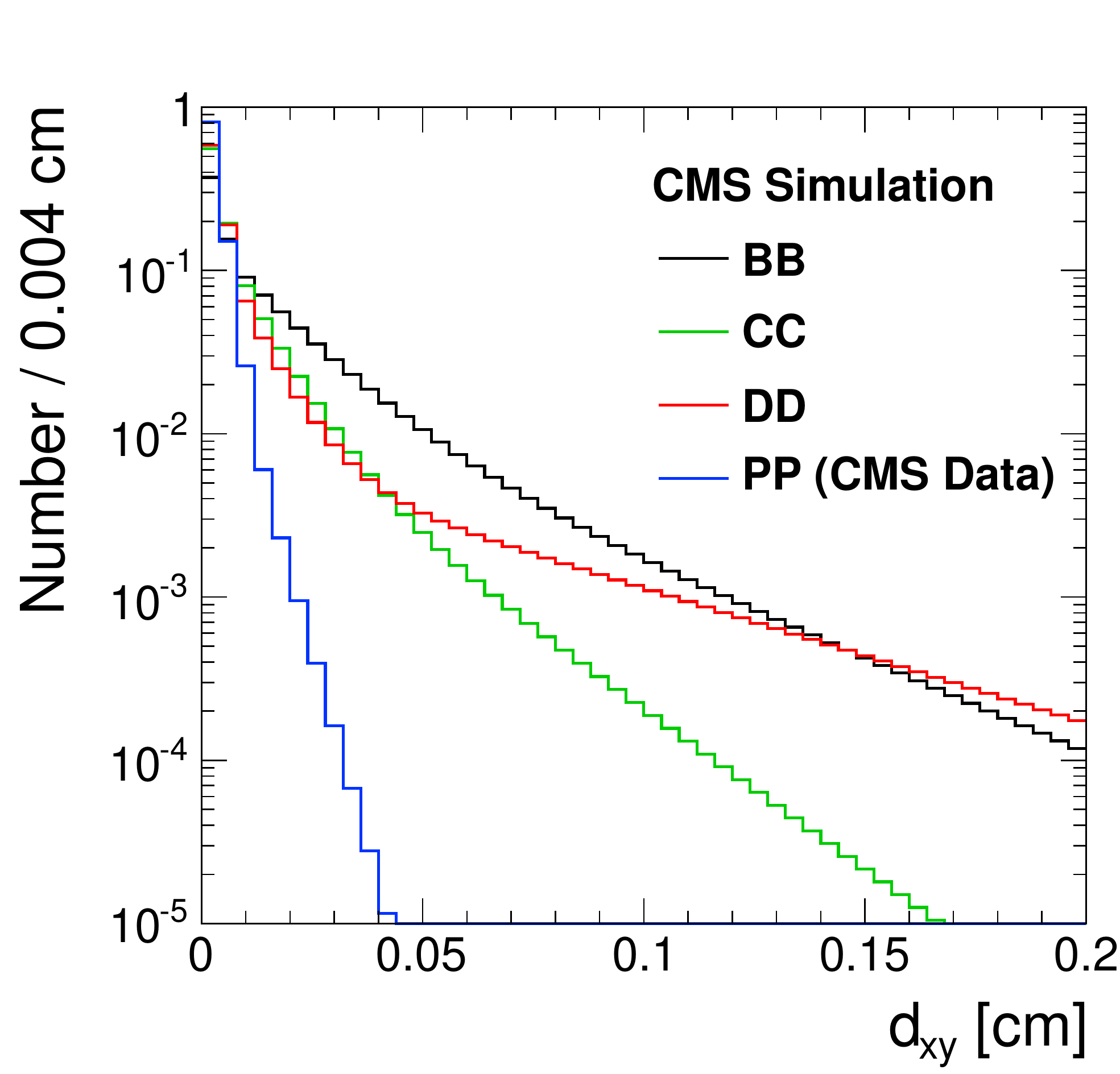}
          \includegraphics[height=0.45\textwidth]{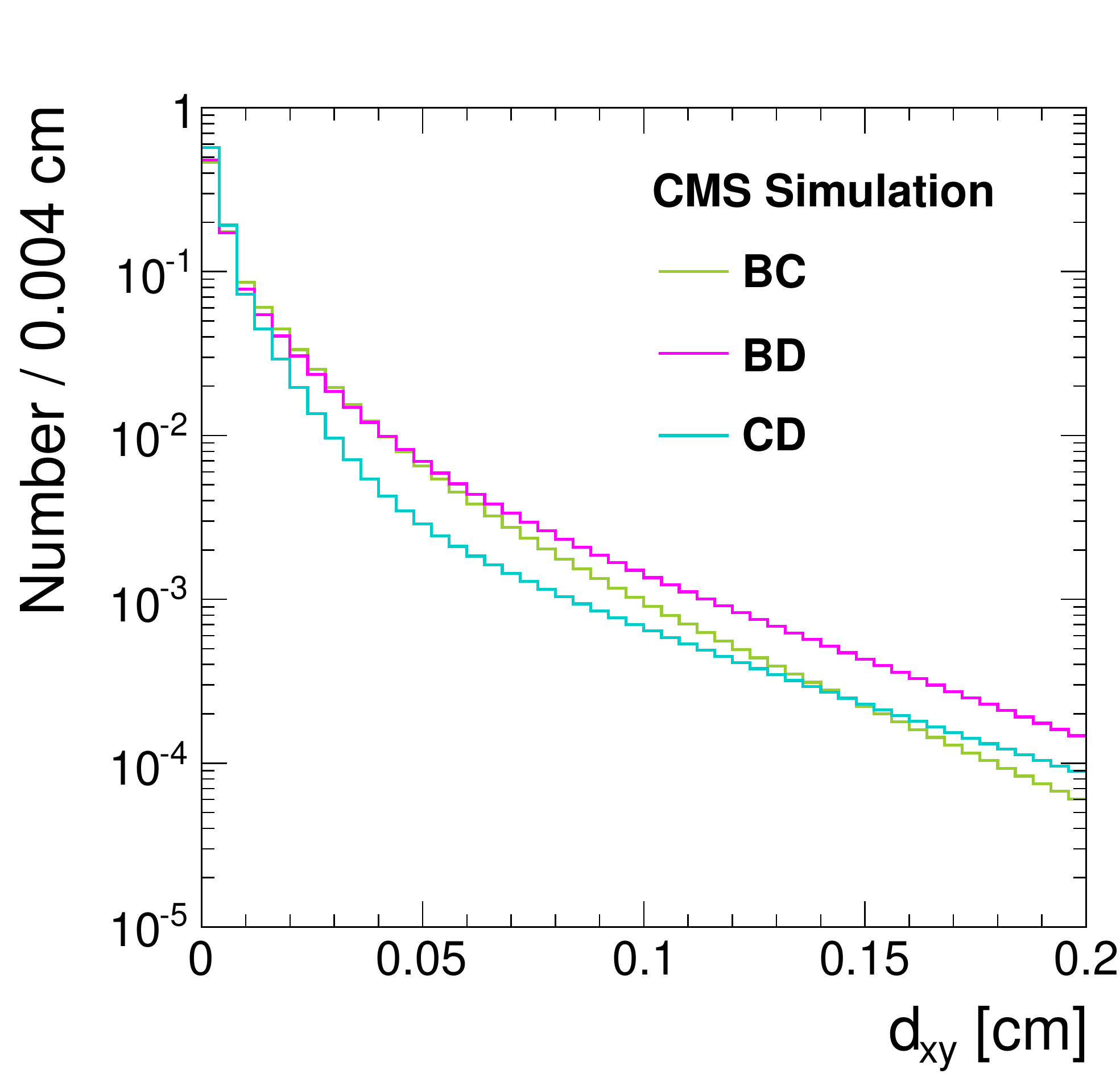}
          \caption{1D projections of the $d_{xy}$ templates used in the fit for muons with $\pt>4\GeV$, for the BB, CC, PP, DD categories (left) and the BC, BD, CD ones (right).}
          \label{f:1Dtemplate_proj}
      \end{center}
    \end{figure}

 The 1D projections of the seven templates
 are shown in Fig.~\ref{f:1Dtemplate_proj}
 for muons with $\pt>4\GeV$.

\section{Measurement of the sample composition} \label{s:fit}
Consistent with the symmetric 2D templates, the data events
are also randomized
by taking the impact parameters of the two muons in each event,
and filling the bin corresponding to
$[d_{xy}(\mu_{1}),\,d_{xy}(\mu_{2})]$ or to $[d_{xy}(\mu_{2}),\,d_{xy}(\mu_{1})]$
according to the outcome of a random number generator.

The fractions of the individual contributions to the
observed distribution are determined with a binned maximum-likelihood
fit.
The fit minimizes the function:

  \begin{linenomath}
  \begin{equation}
   -2
    \mathrm{ln}(L) =  -2\left\{\sum_{i,j=1}^{7}\left[n_{ij} \mathrm{ln}(l_{ij})-l_{ij}\right] - \frac{1}{2}\sum_{k'=1}^{3}\left(\frac{r_{k'}-r^{MC}_{k'}}{\sigma^{MC}_{r_{k'}}}\right)^2\right\},
    \label{eq:likelihood1}
  \end{equation}
  \end{linenomath}

  where $n_{ij}$ is the content
  of the data histogram in the bin ($i,\,j$), $l_{ij}=\sum_{k}[f_k\cdot T_{k,ij}]$, where
$T_k$ is the
 $k^{\text{th}}$ template
  ($k=1,\ldots,7$), and $f_k$ is the fit parameter expressing the fraction
  of events from the $k^{\text{th}}$ source.
The fitted fractions are subject to the normalization
condition $\sum_{k=1}^{7} f_k =1$.
To reduce the number of fit parameters and ease the fit convergence,
the three parameters $f_{BC}, f_{BD},$ and $f_{CD}$ are constrained so
that the ratios $f_{BC}/f_{BB}$, $f_{BD}/f_{BB}$, and $f_{CD}/f_{CC}$
are compatible with the MC expectations
within their statistical uncertainties.
In Eq.~(\ref{eq:likelihood1}), $k'$ is the index of the constrained
templates (BC, BD, CD),
$r_{k'}$ is the ratio of the constrained fit fraction with
respect to the reference fit fraction (for instance in the BC
case $r_{BC}=f_{BC}/f_{BB}$), $r^{MC}_{k'}$ is the ratio of
the constrained fraction and reference fraction in the simulation, and
$\sigma^{MC}_{r_{k'}}$ its statistical uncertainty from the number
of simulated events.

The BC component originates from the production of an extra $\mathrm{c\overline{c}}$
pair from gluon splitting in a $\mathrm{b\overline{b}}$ event.
The production rate of  $\mathrm{c\overline{c}}$
pairs from gluon splitting has been measured at LEP~\cite{bc_fraction,bc_fraction2,bc_fraction3},
and found to be $50\%$ higher than theoretical predictions~\cite{bc_fraction_theory}.
The measured $\mathrm{b\overline{b}}$ rate~\cite{bb_fromgs1,bb_fromgs2,bb_fromgs3} is about
10 times smaller and has a negligible effect on the BC component.
In contrast, the BD and CD contributions are related to the misidentified muon rate in
events with true B and C production. These rates are determined from the
MC simulation, and have been checked using direct measurements
in the data~\cite{bib-muonreco}.
The systematic uncertainties on the fit constraints are discussed in Section~\ref{s:fitsyst}.

\begin{figure}[!htb]
      \begin{center}
          \includegraphics[height=0.45\textwidth]{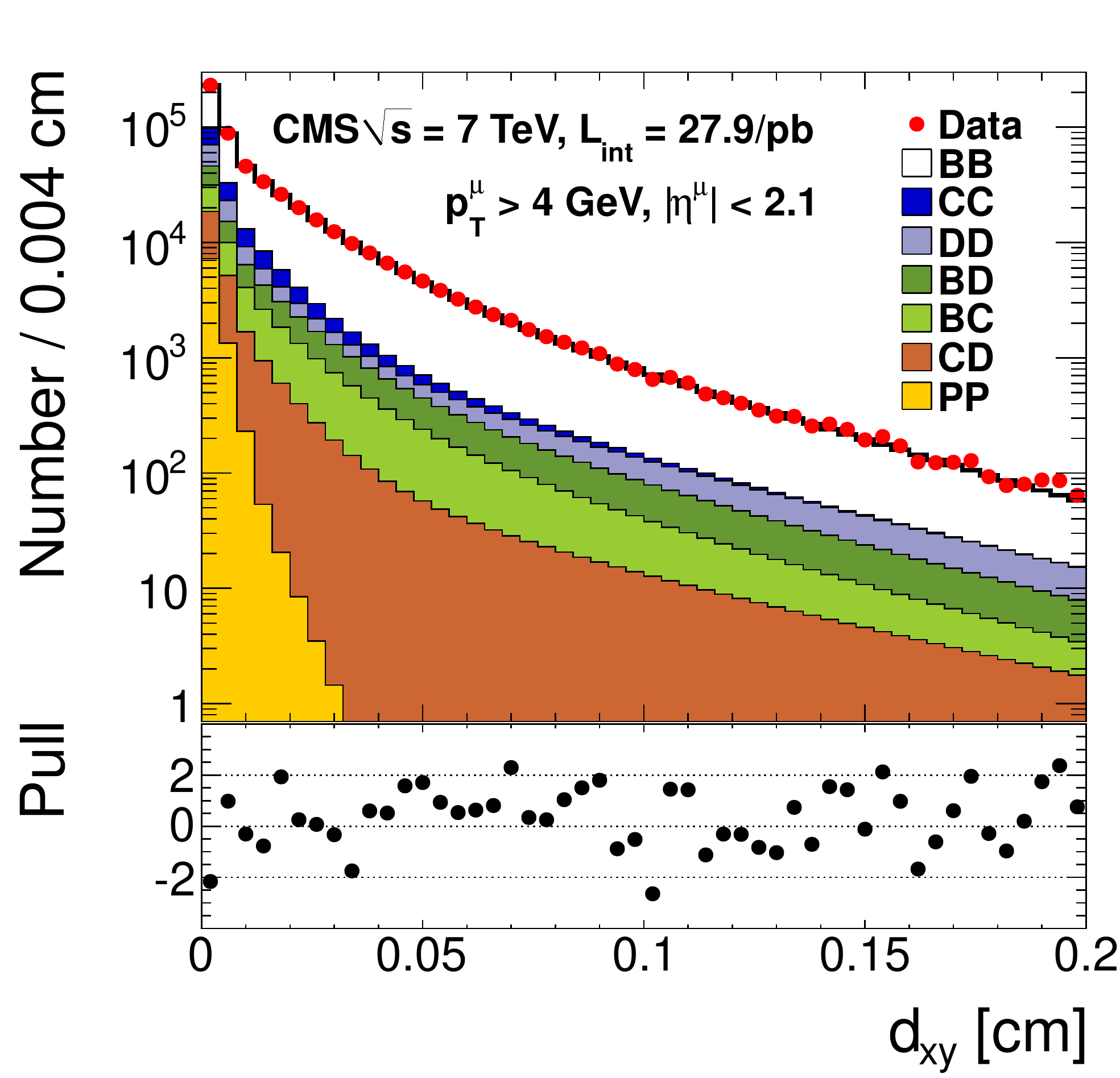}
          \includegraphics[height=0.45\textwidth]{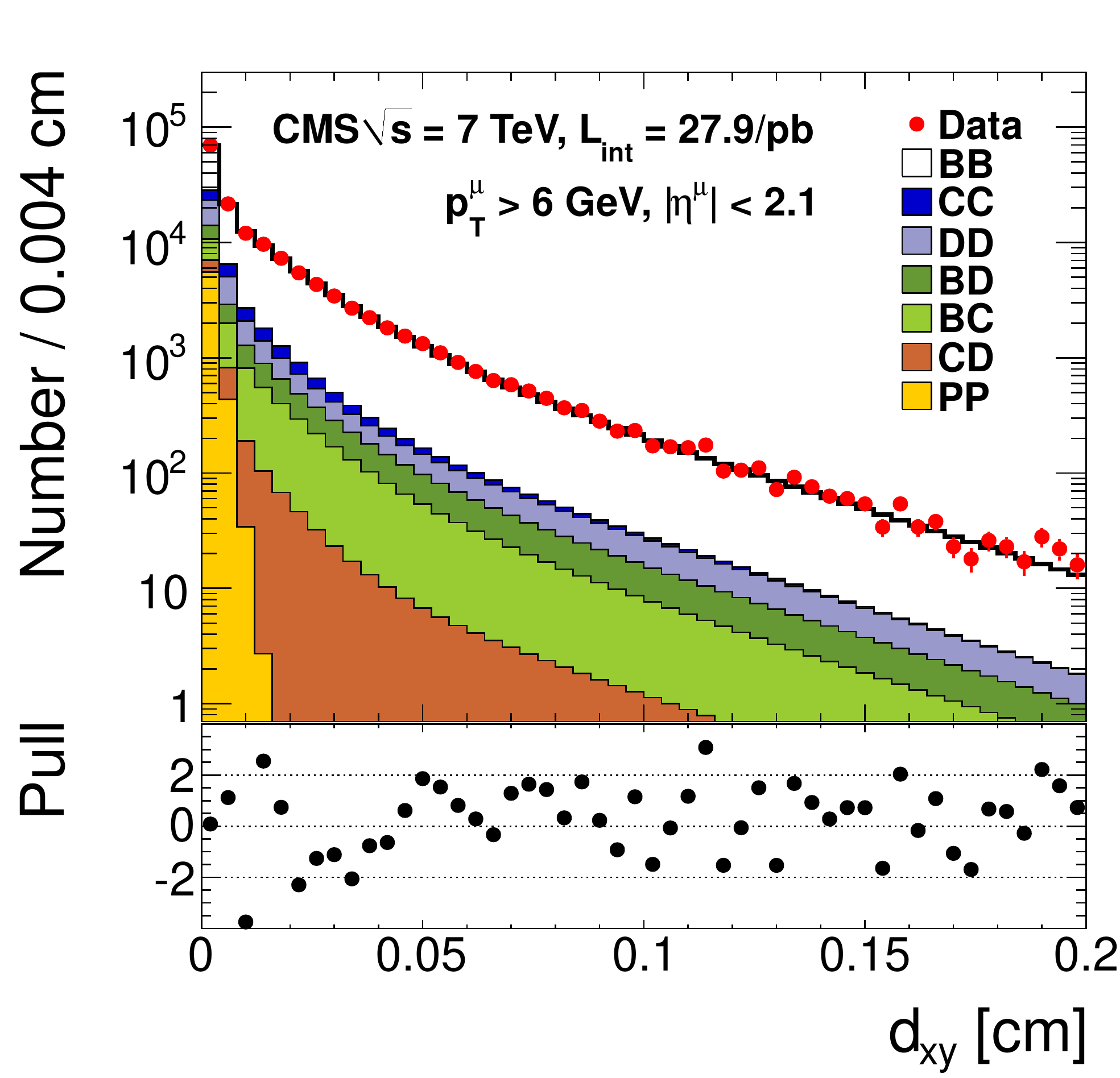}
          \caption{Top: The projected $d_{xy}$ distributions from data with the results
of the fit for muons with $\pt>4\GeV$ (left) and $\pt>6\GeV$ (right). The distribution
from each dimuon source is shown by the histograms. Bottom: The pull distribution
from the fit.}
          \label{f:fitpro}
      \end{center}
    \end{figure}

Table~\ref{tab:bb_fraction_data} gives the results of the fit to the data sample.
The quoted uncertainties are obtained from the fit and are statistical only.
The measured BB fraction is smaller than expected from the simulation, while
the DD fraction is larger.
Projections of the
$d_{xy}$ distributions with the results of the fits
are shown in Fig.~\ref{f:fitpro} for the two $\pt$ selections.

\begin{table}[h!]
\centering
\caption{Results of the likelihood fit to data for the percentage of each dimuon
source with two different muon \pt requirements. The BC, BD, and CD fractions are
      constrained to their ratios to BB and CC fractions as expected from the
      simulation.}
\label{tab:bb_fraction_data}
\begin{tabular}{c|c|c}
\hline
Source         & $\pt>4\GeV$           & $\pt>6\GeV$          \\
\hline
BB             & $66.8 \pm 0.3$      & $70.2 \pm 0.3$     \\
CC             & $9.2 \pm 0.6$       & $5.5 \pm 1.2$      \\
BC             & $5.2 \pm 0.1$       & $4.9 \pm 0.1$      \\
PP             & $1.7 \pm 0.3$       & $4.0 \pm 0.4$      \\
DD             & $7.8 \pm 1.1$       & $9.5 \pm 2.1$      \\
BD             & $5.6 \pm 0.1$       & $4.2 \pm 0.1$      \\
CD             & $3.7 \pm 0.9$       & $1.6 \pm 0.5$      \\
\hline
\end{tabular}
\end{table}

\section{Efficiency determination} \label{s:eff}

The total efficiency $\epsilon$ is defined as the fraction of signal events
produced within the acceptance ($\pt>4\GeV$ or $\pt>6\GeV$,
$|\eta| < 2.1$ for each muon) that are retained in the analysis. In
the simulation, the values of
 $\epsilon_{\mathrm{MC}} = (44.3 \pm 0.1) \%$ and $(69.9
 \pm 0.1) \%$ are computed for signal events with a $\pt$ threshold of 4
 and 6\GeV, respectively.

To compare these values to efficiencies measured in data,
the selection procedure is divided into three steps,
each defined relative to events passing the previous one:

\begin{enumerate}
 \item muon selection (``MuSel''):  events having at least two
   selected muons, each associated with a reconstructed vertex;
 \item event selection (``EvSel''):  events passing the dimuon invariant mass requirements, with both muons belonging to the same vertex;
 \item trigger selection (``Trg''):  events passing the trigger requirements.
\end{enumerate}

The efficiencies obtained by counting the signal events passing each step in the simulation are given
in Table \ref{tab:eff_evCounting}.

The total efficiency can alternatively be expressed on an event-by-event basis by
defining the efficiency $\epsilon_i$ to select the $i^{\mathrm{th}}$ signal event as
$\epsilon_i = \epsilon_{i,\mathrm{MuSel}} \cdot  \epsilon_{i, \mathrm{EvSel}} \cdot \epsilon_{i,\mathrm{Trg}}$.
The $(\pt,\eta)$ distribution of the signal
events and the efficiency $\epsilon_{i, \text{EvSel}}$ can only be extracted from simulation.
The efficiencies $\epsilon_{i,\text{MuSel}}$ and $\epsilon_{i,\text{Trg}}$ can be found
as the products of the single-muon efficiencies,
$\epsilon_{i}=\epsilon_{\mu_1}(\pt,\eta)\cdot \epsilon_{\mu_2}(\pt,\eta)$,
under the assumption that the single-muon efficiencies $\epsilon_{\mu_i}$
only depend on the $\pt$ and $\eta$ of the muon.
This assumption is found to be compatible with
the efficiencies determined in the simulated sample, within their statistical uncertainties.

\begin{table}[h!]
\centering
\caption{ Efficiencies (in percent) at each step of the analysis found from the
  simulation and from the data. The last
  column reports the overall efficiency, obtained from the product of
  the three efficiencies shown. The event selection efficiency $\epsilon_{\text{EvSel}}$
 cannot be found with the data, so the MC simulation value is used.
 The bias and feed-through corrections
  described in the text are also included in the overall efficiency.
 Only statistical uncertainties are reported.}
\label{tab:eff_evCounting}
\begin{tabular}{c||c|c|c||c}
\hline
  Sample &  $\epsilon_{\text{MuSel}}$        &  $\epsilon_{\text{EvSel}}$      & $\epsilon_{\text{Trg}}$    &$\epsilon$ \\
\hline
\hline
$\pt > 4\GeV$, MC  &    $64.8\pm0.1$     &   $78.0\pm0.1$   & $87.7\pm0.1$    & $44.3\pm0.1$    \\
$\pt > 4\GeV$, data  &    $69.5\pm3.6$     &   -   & $86.1\pm2.0$    & $48.8\pm2.9$    \\
\hline
$\pt > 6\GeV$, MC  &    $83.6\pm0.1$     &   $90.1\pm0.1$   & $92.8\pm0.1$    & $69.9\pm0.1$    \\
$\pt > 6\GeV$, data  &    $87.0\pm3.4$     &   -   & $93.4\pm2.1$    & $74.4\pm3.8$    \\
\hline
\end{tabular}
\end{table}

In the data, the single-muon selection and trigger efficiencies are measured in
intervals of $\pt$ and $\eta$ with the
``tag-and-probe'' (T\&P) method~\cite{bib-muonreco,bib-trackingefficiency},
which employs a sample of $\JPsi\rightarrow\Pgmp\Pgmm$ events
selected with minimal trigger requirements.
The selection efficiency found from this method is consistent with the value from the simulation
($\epsilon_{\text{MuSel}}^{\text{data}} / \epsilon_{\text{MuSel}}^{\mathrm{MC}} =
1.073 \pm 0.054$ for $\pt>4\GeV$ and $1.041\pm0.047$ for $\pt>6\GeV$), as is
the trigger efficiency
($\epsilon_{\text{Trg}}^{\text{data}} / \epsilon_{\text{Trg}}^{\mathrm{MC}} =
0.982 \pm 0.028$ for $\pt>4\GeV$ and $1.006\pm0.023$ for $\pt>6\GeV$).

Differences in the kinematic distributions between the $\JPsi$ sample and the $\cPqb\cPaqb$ events
might imply different bin-averaged efficiencies, causing biases in the
region close to the acceptance thresholds.
An overall bias correction of $0.966 \pm 0.015$ ($1.004\pm0.012$) is computed
when comparing the efficiencies in the simulation computed with the T\&P method and
those obtained with the signal in the range
$\pt>4\GeV$ ($\pt>6\GeV$), where the uncertainties are statistical only.

Another correction to the total efficiency is applied to take into account the
feed-through of events
where one of the muons has a true \pt below the selection limit,
whereas the reconstructed \pt is above it.
This effect is computed using the simulation
by finding the fraction of selected events with at least one muon generated
outside of the acceptance, and is equal to $0.990$ ($0.980$) for $\pt>4\,(6)\GeV$,
with negligible uncertainties.

The overall efficiency
is computed as the product of the
efficiencies for the muon selection and trigger, as obtained with
the T\&P method in data, times the event selection efficiency found in the simulation,
divided by the bias and the
feed-through corrections. Results are shown in Table~\ref{tab:eff_evCounting}.

\section{Systematic uncertainties}
 \label{s:syst}
Several sources of systematic uncertainties have been considered for
this measurement. They are divided into uncertainties due to the
model dependencies for both the signal and the backgrounds,
the effects related to the impact parameter resolution,
the fit method, and the
measurement of the efficiency.
Each of these is described separately in the subsections below.

\subsection{Model-dependent uncertainties}

The impact parameter projected onto the
plane transverse to the beam axis
of a muon produced in a hadron semileptonic
decay is
related
to the parent hadron's proper decay time $t$ by:

\begin{equation}
d_{xy} = ~\beta \gamma ~ct ~\mathrm{sin} \delta ~\mathrm{sin} \theta,
\end{equation}

where $\beta$ is the ratio of the
hadron
velocity and the speed of
light $c$, $\gamma = (1-\beta^2)^{1/2}$, $\delta$ is the angle between the
muon and the hadron directions in the laboratory frame,
and $\theta$ is the polar angle between the hadron direction and the beam
axis. Uncertainties in the parent lifetime affect the simulation of the proper distance
distribution $ct$, and uncertainties in the parent hadron energy spectrum affect the Lorentz
boost factor
$\beta\gamma$ and the angle $\delta$. The three general
categories of systematic uncertainties due to these model dependencies are:

{\it b- and c-hadron properties:} four of the long-lived $\PB$ hadrons
produced at the LHC decay to muons at a non-negligible rate.
While the $\PB_\mathrm{d}$ and $\PB_\mathrm{u}$
lifetimes are known with a precision better than 1\%, the
$\PB_\mathrm{s}$ and $\PgL_\mathrm{b}$ lifetimes are 
measured with larger uncertainties.
Simulated MC events with $\PB_\mathrm{s}$
and $\PgL_\mathrm{b}$ decays are reweighted so as to vary the
corresponding lifetimes by their uncertainties
\cite{HFAG2}, the templates are recomputed, and the fit is repeated.
The fit result changes by $\pm 2.1\%$ ($\pm1.5\%$) for $\pt>4\,(6)\GeV$.
The effects from uncertainties on the $\PB_\mathrm{d},
\PB_\mathrm{u}$, and $\mathrm{c}$-hadron lifetimes, similarly evaluated, are negligible.
The $\mathrm{b}$-hadron sample composition has been measured by experiments at LEP and the Tevatron~\cite{HFAG2}, 
and by LHCb~\cite{LHCbLambdab}.
While substantial agreement has been found for the $\PB_\mathrm{s}/\left(\PB_\mathrm{u}+\PB_\mathrm{d}\right)$
fraction, a sizable discrepancy was observed for
$f_{\PgL_\mathrm{b}}=\PgL_\mathrm{b}/\left(\PB_\mathrm{u}+\PB_\mathrm{d}\right)$.
The results presented here are obtained using the averages between the LEP
and LHCb results, $f_{\PgL_\mathrm{b}}=0.18\pm0.09$ ($0.165\pm0.075$) for $\pt>4\,(6)\GeV$,
where the uncertainties correspond to half the difference between LEP and LHCb.
Varying $f_{\PgL_\mathrm{b}}$ in these ranges affects the measurement by $\pm2.7\%$ ($\pm1.8\%$)
for $\pt>4\,(6)\GeV$. Varying the other parameters affecting the
$\mathrm{b}$-hadron and $\mathrm{c}$-hadron sample compositions
by their uncertainties
has a smaller effect for both $\pt>4~\GeV$ and $\pt>6~\GeV$
($\pm 0.7\%,\,\pm 0.8\%$, respectively).

\textit{$b$-quark properties:} uncertainties in the production of
 $\PB$ hadrons from the fragmentation of a $\mathrm{b}$ quark affect both the
 shape of the $d_{xy}$ distribution and the efficiency estimate.
 The systematic uncertainty is computed as the difference between the
 default
result and those obtained with two different hadronization
 models in the \PYTHIA simulation:
 the Lund symmetric \cite{pythia64} and the Peterson \cite{peterson}
 functions.
Taking into account the effects on the $\mathrm{b}$ templates and those
 connected with the extraction of the efficiency, overall
 uncertainties of $\pm 3.3\%$ ($\pt>4\GeV$) and $\pm 3.6\%$ ($\pt>6\GeV$) are obtained.
Using different PDFs to describe $\cPqb$-quark production in $\Pp\Pp$
collisions has an effect of of $\pm 0.9\%$ ($\pt>4\GeV$) and $\pm 0.5\%$ ($\pt>6\GeV$).

{\it Light-meson decays in flight:} muons from $\Pgp$ and $\PK$
 decays have different $d_{xy}$ distributions. The shape is also
 different
 for muons from light mesons
 produced in the hadronization of a
 light quark, or from the decay of a heavy hadron.
Given the uncertainties on the pion and kaon fractions in the simulation,
we vary the relative amounts by $\pm 30\%$ and find a negligible effect on the
final results.  Similarly, we change the ratio of light mesons from
heavy-flavor and light-flavor decays
by $\pm 50\%$, and observe a $2.5\%$ ($2.6\%$) change in the results for $\pt>4\,(6)\GeV$.
 The generator-level filter applied to the simulated sample, requiring two
 muons to be produced within the tracker volume in each event,
 affects the shape and composition of the decays-in-flight template.
The impact of the filter on the BB fraction is estimated by
extracting the decays-in-flight template from an unbiased simulated sample
in which only one generated muon is required to pass the filter and the
other muon is used in determining the template. Repeating the analysis
with this new template results in a $0.5\%$ variation of the final result
for both $\pt$ selections.

The total model-dependent systematic uncertainty, found by adding in
quadrature the contributions listed above, is $5.5\%$ for $\pt>4\GeV$
and $5.1\%$ for $\pt>6\GeV$.

\subsection{Uncertainties on the impact parameter resolution}

The systematic uncertainty from the impact
parameter resolution is determined
by comparing the $d_{xy}$ distribution from
 prompt $\PgUa \rightarrow \Pgmp\Pgmm$ decay candidates reconstructed in collision data
to the predicted distribution from MC
 simulation.
 A slight $\phi$ dependence in the
 determination of the signed impact parameter with respect to
the beam spot, where $\phi$ is the azimuthal angle of the muon track,
  due to the CMS tracker not being
perfectly centered around the beam pipe, is not reproduced by the simulation. The combined effect
from the misalignment and the different
resolution in data and simulation is evaluated
by an additional smearing of the impact parameter consistent with the
observed differences between data and simulation.
A further check to avoid the region dominated by the resolution
has been performed by moving the lower bound of the fit range from 0 to 40\,$\mu$m.
The maximum deviation from the default result found with these two methods
is $2.7\%$ ($4.0\%$) for $\pt>4\,(6)\GeV$,
which is taken as the systematic uncertainty due to the detector resolution.

\subsection{Uncertainties related to the Monte Carlo precision and the fit method} \label{s:fitsyst}

There are four general categories of systematic uncertainty caused
by the MC statistical precision and the fitting procedure.  These
include:

{\it Monte Carlo precision:}
the likelihood fit is validated using a set of 500 parameterized simulated datasets,
each with the same number of events as the data sample.
The fit results from these datasets reproduce the input values
with uncertainties consistent with those obtained in data,
and the pull distribution is well described by a normal function.
The r.m.s.\ of the results obtained for the BB fraction is $0.3\%$ ($0.7\%$) for $\pt>4\,(6)\GeV$,
which are taken as the systematic uncertainties related to the finite simulated sample.

{\it Template parameterization:} the $d_{xy}$ distributions in the
simulated data used for the fit are smoothed
using a
superposition
of a Gaussian plus
one or two
exponential functions,
depending on the extent
of the tail. The associated systematic uncertainty,
evaluated by using different parametrizations,
is equal to $\pm 0.7\%$ for both $\pt$ selections.
The systematic uncertainty from the use of symmetrized templates
is estimated to be $\pm 0.6\%$ ($\pm 0.7\%$) for $\pt>4\,(6)\GeV$,
by comparing the results obtained in the simulation
when a sum of symmetrized templates is used as pseudo-data instead of the usual randomized distribution.

{\it Bin size and fit upper bound:} varying the $d_{xy}$ bin size in the range $0.002 - 0.008\,$cm
accounts for a systematic uncertainty of $1.0\%$ ($2.1\%$) for $\pt>4\,(6)\GeV$,
while varying the fit upper bound in the range $0.15 - 0.25\,$cm accounts
for $0.3\%$ ($0.4\%$).

{\it Fit constraints:} the BC, BD, and CD fractions are constrained in the fit
so that their ratios with respect to the fitted BB fraction in the BC and
BD cases, and the fitted CC fraction in the CD case, agree with the
predicted values from the MC simulation, as described in Section~\ref{s:fit}.
The uncertainties from this procedure include those on the rate of
$\mathrm{c\overline{c}}$ production from gluon splitting and the muon
misidentification rates in the simulation. To estimate the uncertainty,
we vary the constraints on the fractions by $\pm50\%$
around the simulation
values, which
induces a difference of $1.6\%$ ($1.2\%$) for $\pt>4\,(6)\GeV$
in the fitted BB fraction.
Since the 2D fit neglects the
mixing of prompt and non-prompt muon components (PB, PC, PD), an additional
systematic uncertainty is computed
by assigning to the BB fraction an uncertainty equal to
the missing contributions, as found in the simulation, of
$0.7\%$ ($0.6\%$) for $\pt>4\,(6)\GeV$.

The total systematic uncertainty related to the fit method is
found by adding the contributions in quadrature, which gives
$2.2\%$ ($2.7\%$) for $\pt>4\,(6)\GeV$.

As a consistency check, an unconstrained 1D fit
is performed on the $d_{xy}$ distribution of the muons selected for the analysis, using the templates
derived in Section ~\ref{s:muon_classes}.
The results are in agreement within the quoted
systematic uncertainty with those from the 2D fit.

\subsection{Efficiencies from data and the dimuon invariant mass extrapolation}

The statistical uncertainties of the efficiencies found from the
T\&P
method
of $6.0\%$ ($5.2\%$) for $\pt>4\,(6)\GeV$
are taken as the systematic uncertainty
on this procedure.

The dimuon invariant mass distribution predicted from the MC
simulation, scaled to the fitted fractions in the data, does not agree
with the observed distribution within the uncertainties.  Attributing the
entire difference as being due to extra $\bbbar$ signal events, gives us
the largest systematic uncertainty from this source of $1.1\%$ ($3.3\%$) for
$\pt>4\,(6)\GeV$.

\subsection{Overall systematic uncertainty}
All the systematic uncertainties described so far are summarized in Table \ref{t:sys}
and sum in quadrature to $8.9\%$
($9.4\%$) for $\pt>4\,(6)\GeV$, with the larger contribution coming from the data-driven efficiency determination with the T\&P method.
The last source of systematic uncertainty to be considered is related to the integrated
luminosity of the dimuon data sample,
which is determined with a $4\%$ uncertainty~\cite{lumipas}.
The total systematic uncertainty is therefore $9.8\%$ for $\pt>4\GeV$ and $10.2\%$ for $\pt>6\GeV$.

\begin{table}[h!]
  \centering
	\caption{Systematic uncertainties on the cross-section measurements in percent for the two \pt limits.}
  \begin{tabular}{c|c|c}
    \hline
    Source & \multicolumn{2}{c}{ Uncertainty } \\
    \hline
                                           & $\pt>4\GeV$  & $\pt>6\GeV$ \\
    \hline

    Model dependency                        &  $ 5.5$  &  $ 5.1$ \\
    Impact parameter resolution             &  $ 2.7$  &  $ 4.0$  \\
    Monte Carlo precision and fit method    &  $ 2.2$  &  $ 2.7$  \\
    Efficiencies and acceptance             &  $ 6.1$  &  $ 6.2$  \\

    \hline
    Total                                   & $ 8.9$   & $ 9.4$  \\

\end{tabular}

\label{t:sys}
\end{table}

\section{Results and comparison with QCD predictions} \label{s:cross_section}

The $\Pp\Pp\rightarrow \bbbar \mbox{X}\rightarrow \Pgm\Pgm \mbox{X}^\prime$ cross section within the accepted kinematic range is determined from the
observed number of dimuon events passing the event selection ${\cal N}_{\mu\mu}$, the fraction of signal events in
the dimuon sample $f_{BB}$, the average efficiency
for the trigger, muon identification, and event selection $\epsilon$,
weighted by the \pt and $\eta$ distributions,
and the integrated luminosity $\mathcal{L}$ according to the
relation:
\begin{linenomath}
\begin{equation}
\sigma ( \Pp\Pp \rightarrow \bbbar \mbox{X} \rightarrow \Pgm\Pgm \mbox{X}^\prime, \pt>4\mathrm{\ or\ 6\GeV}, |\eta|<2.1) =
  \frac {{\cal N}_{\mu\mu} \cdot f_{BB}} {\epsilon \cdot {\cal L}}.
\label{eq:xsection}
\end{equation}
\end{linenomath}

By applying Eq.~(\ref{eq:xsection}) we measure:
\begin{linenomath}
\begin{eqnarray}
\lefteqn{\sigma (\Pp\Pp\rightarrow \bbbar \mbox{X} \rightarrow \Pgm\Pgm \mbox{X}^\prime,\pt>4\GeV,|\eta|<2.1) = } \hspace{1.25in}\\ \nonumber
& & 26.4\pm0.1 \mbox{ (stat.) }\pm2.4\mbox{ (syst.) }\pm1.1\mbox{ (lumi.) }\,\mathrm{nb}
\end{eqnarray}
\end{linenomath}
and
\begin{linenomath}
\begin{eqnarray}
\lefteqn{\sigma (\Pp\Pp\rightarrow \bbbar \mbox{X} \rightarrow \Pgm\Pgm \mbox{X}^\prime,\pt>6\GeV,|\eta|<2.1) = } \hspace{1.25in} \\ \nonumber
& & 5.12\pm0.03 \mbox{ (stat.) }\pm0.48\mbox{ (syst.) }\pm0.20\mbox{ (lumi.) }\unit{nb}.
\end{eqnarray}
\end{linenomath}

The
cross sections predicted by the leading-order \PYTHIA simulation
are
48.2\unit{nb} for $\pt>4\GeV$ and 9.2\unit{nb} for $\pt>6\GeV$,
where the statistical uncertainties are negligible.
That \PYTHIA predicts a cross section value higher than the one measured in data has
been seen in previous analyses~\cite{BPH_10_007},
and is confirmed by our present findings.

The next-to-leading-order event generator \MCATNLO~\cite{mc_at_nlo} is used
to estimate the NLO QCD prediction for this measurement,
with the CTEQ6.6
PDF
and a $\mathrm{b}$-quark mass of $4.75\GeV$.
The generator is interfaced with \HERWIG~\cite{herwig} for parton showering, hadronization, and decays.
The systematic
uncertainty for this prediction is obtained by varying the $\cPqb$-quark
mass between 4.5\GeV and 5\GeV,
and by changing the
PDF to the MSTW2008~\cite{MSTW2008} set.
The scale uncertainty is estimated by varying the QCD renormalization and
factorization scales
independently from half to twice their default values,
as in Ref.~\cite{Cacciari:2003uh}.

The predicted cross sections are:
 \begin{linenomath}
 \begin{equation}
   \sigma_{\MCATNLO} (\Pp\Pp\rightarrow \bbbar \mbox{X} \rightarrow \Pgm\Pgm \mbox{X}^\prime,\pt>4\GeV,|\eta|<2.1)=
   19.7\pm 0.3
\mbox{ (stat.) } 
 {}^{+6.5}_{-4.1}\mbox{ (syst.) } \mbox{ nb}
 \end{equation}
 \end{linenomath}
and
 \begin{linenomath}
 \begin{equation}
   \sigma_{\MCATNLO} (\Pp\Pp\rightarrow \bbbar \mbox{X} \rightarrow \Pgm\Pgm \mbox{X}^\prime,\pt>6\GeV,|\eta|<2.1)=
   4.40\pm 0.14
\mbox{ (stat.) } 
 {}^{+1.10}_{-0.84}\mbox{ (syst.) } \mbox{ nb}.
 \end{equation}
 \end{linenomath}

Both predictions are compatible with our results within the uncertainties of the NLO calculations and the measurements.

\section{Summary}\label{s:conc}

A measurement of the inclusive cross section for the process \mbox{$ \Pp\Pp\rightarrow \bbbar \mbox{X}
\rightarrow \Pgm\Pgm \mbox{X}^\prime$}  at $\sqrt{s} = 7\TeV$ has been presented,
based on an integrated luminosity of $27.9\pm 1.1$\,pb$^{-1}$ collected by the
CMS experiment at the LHC. Selecting pairs of muons each with
pseudorapidity $|\eta|<2.1$, the value
$\sigma (\Pp\Pp\rightarrow \bbbar \mbox{X}
\rightarrow \Pgm\Pgm \mbox{X}^\prime)=26.4\pm0.1 \mbox{ (stat.) }\pm2.4\mbox{ (syst.) }\pm1.1\mbox{ (lumi.) }\,$nb
was obtained for muons with transverse momentum $\pt>4\GeV$, and
$5.12\pm 0.03\mbox{ (stat.) }\pm0.48\mbox{ (syst.) }\pm0.20\mbox{ (lumi.) }$\unit{nb}
for muons with $\pt>6\GeV$. 
 This result is the most precise measurement of this quantity yet made at the LHC.

\section*{Acknowledgements}
\label{sec:ack}

\hyphenation{Bundes-ministerium Forschungs-gemeinschaft Forschungs-zentren} We congratulate our colleagues in the CERN accelerator departments for the excellent performance of the LHC machine. We thank the technical and administrative staff at CERN and other CMS institutes. This work was supported by the Austrian Federal Ministry of Science and Research; the Belgium Fonds de la Recherche Scientifique, and Fonds voor Wetenschappelijk Onderzoek; the Brazilian Funding Agencies (CNPq, CAPES, FAPERJ, and FAPESP); the Bulgarian Ministry of Education and Science; CERN; the Chinese Academy of Sciences, Ministry of Science and Technology, and National Natural Science Foundation of China; the Colombian Funding Agency (COLCIENCIAS); the Croatian Ministry of Science, Education and Sport; the Research Promotion Foundation, Cyprus; the Ministry of Education and Research, Recurrent financing contract SF0690030s09 and European Regional Development Fund, Estonia; the Academy of Finland, Finnish Ministry of Education and Culture, and Helsinki Institute of Physics; the Institut National de Physique Nucl\'eaire et de Physique des Particules~/~CNRS, and Commissariat \`a l'\'Energie Atomique et aux \'Energies Alternatives~/~CEA, France; the Bundesministerium f\"ur Bildung und Forschung, Deutsche Forschungsgemeinschaft, and Helmholtz-Gemeinschaft Deutscher Forschungszentren, Germany; the General Secretariat for Research and Technology, Greece; the National Scientific Research Foundation, and National Office for Research and Technology, Hungary; the Department of Atomic Energy and the Department of Science and Technology, India; the Institute for Studies in Theoretical Physics and Mathematics, Iran; the Science Foundation, Ireland; the Istituto Nazionale di Fisica Nucleare, Italy; the Korean Ministry of Education, Science and Technology and the World Class University program of NRF, Korea; the Lithuanian Academy of Sciences; the Mexican Funding Agencies (CINVESTAV, CONACYT, SEP, and UASLP-FAI); the Ministry of Science and Innovation, New Zealand; the Pakistan Atomic Energy Commission; the Ministry of Science and Higher Education and the National Science Centre, Poland; the Funda\c{c}\~ao para a Ci\^encia e a Tecnologia, Portugal; JINR (Armenia, Belarus, Georgia, Ukraine, Uzbekistan); the Ministry of Education and Science of the Russian Federation, the Federal Agency of Atomic Energy of the Russian Federation, Russian Academy of Sciences, and the Russian Foundation for Basic Research; the Ministry of Science and Technological Development of Serbia; the Ministerio de Ciencia e Innovaci\'on, and Programa Consolider-Ingenio 2010, Spain; the Swiss Funding Agencies (ETH Board, ETH Zurich, PSI, SNF, UniZH, Canton Zurich, and SER); the National Science Council, Taipei; the Scientific and Technical Research Council of Turkey, and Turkish Atomic Energy Authority; the Science and Technology Facilities Council, UK; the US Department of Energy, and the US National Science Foundation.

Individuals have received support from the Marie-Curie programme and the European Research Council (European Union); the Leventis Foundation; the A. P. Sloan Foundation; the Alexander von Humboldt Foundation; the Belgian Federal Science Policy Office; the Fonds pour la Formation \`a la Recherche dans l'Industrie et dans l'Agriculture (FRIA-Belgium); the Agentschap voor Innovatie door Wetenschap en Technologie (IWT-Belgium); the Council of Science and Industrial Research, India; and the HOMING PLUS programme of Foundation for Polish Science, cofinanced from European Union, Regional Development Fund. 

\bibliography{auto_generated}   

\cleardoublepage \appendix\section{The CMS Collaboration \label{app:collab}}\begin{sloppypar}\hyphenpenalty=5000\widowpenalty=500\clubpenalty=5000\textbf{Yerevan Physics Institute,  Yerevan,  Armenia}\\*[0pt]
S.~Chatrchyan, V.~Khachatryan, A.M.~Sirunyan, A.~Tumasyan
\vskip\cmsinstskip
\textbf{Institut f\"{u}r Hochenergiephysik der OeAW,  Wien,  Austria}\\*[0pt]
W.~Adam, T.~Bergauer, M.~Dragicevic, J.~Er\"{o}, C.~Fabjan, M.~Friedl, R.~Fr\"{u}hwirth, V.M.~Ghete, J.~Hammer\cmsAuthorMark{1}, M.~Hoch, N.~H\"{o}rmann, J.~Hrubec, M.~Jeitler, W.~Kiesenhofer, M.~Krammer, D.~Liko, I.~Mikulec, M.~Pernicka$^{\textrm{\dag}}$, B.~Rahbaran, C.~Rohringer, H.~Rohringer, R.~Sch\"{o}fbeck, J.~Strauss, A.~Taurok, F.~Teischinger, P.~Wagner, W.~Waltenberger, G.~Walzel, E.~Widl, C.-E.~Wulz
% ...
\vskip\cmsinstskip
\dag:~Deceased\\
1:~~Also at CERN, European Organization for Nuclear Research, Geneva, Switzerland\\
% ...

\end{sloppypar}
\end{document}